\documentclass[manuscript]{aastex631}

\definecolor{Red}{rgb}{1,0,0}

\usepackage{bm}

\received{} 
\revised{} 
\accepted{} 
\submitjournal{ApJ}

\shorttitle{Ion scale waves and dense ion beams}
\shortauthors{Yogesh et al.}
\begin{document}

\title{Evidence of Interaction between Ion-Scale Waves and Ion Velocity Distributions in the Solar Wind}

\author[0000-0001-6018-9018]{Yogesh}
\affiliation{The Catholic University of America, Washington, DC 20064, USA}
\affiliation{Heliophysics Science Division, NASA Goddard Space Flight Center, 
Greenbelt, MD 20771, USA}

\author[0000-0003-0602-6693]{Leon Ofman}
\affiliation{The Catholic University of America, Washington, DC 20064, USA}
\affiliation{Heliophysics Science Division, NASA Goddard Space Flight Center, 
Greenbelt, MD 20771, USA}
\affiliation{Visiting, Tel Aviv University, Tel Aviv, Israel}

\author[0000-0002-5240-044X]{Scott A Boardsen}
\affiliation{Goddard Planetary Heliophysics Institute, University of Maryland, Baltimore County, MD 21250, USA}
\affiliation{Heliophysics Science Division, NASA Goddard Space Flight Center, 
Greenbelt, MD 20771, USA}

\author[0000-0001-6038-1923]{Kristopher Klein}
\affiliation{Lunar and Planetary Laboratory, University of Arizona, Tucson, AZ 85721, USA}

\author[0000-0002-7365-0472]{Mihailo Martinovi\'{c}}
\affiliation{Lunar and Planetary Laboratory, University of Arizona, Tucson, AZ 85721, USA}

\author[0000-0002-4001-1295]{Viacheslav M Sadykov}
\affiliation{Physics \& Astronomy Department, 
Georgia State University, 
Atlanta, GA 30303, USA}

\author[0000-0003-1138-652X]{Jaye L Verniero}
\affiliation{Heliophysics Science Division,
NASA Goddard Space Flight Center,
Greenbelt, MD 20771, USA}

\author[0000-0002-8941-3463]{Niranjana Shankarappa}
\affiliation{Lunar and Planetary Laboratory, University of Arizona, Tucson, AZ 85721, USA}

\author[0000-0002-6849-5527]{Lan K Jian}
\affiliation{Heliophysics Science Division, 
NASA Goddard Space Flight Center, 
Greenbelt, MD 20771, USA}

\author[0000-0002-3808-3580]{Parisa Mostafavi}
\affiliation{Johns Hopkins Applied Physics Laboratory, 
Laurel, MD 20723, USA}

\author[0000-0002-9954-4707]{Jia Huang}
\affiliation{Space Sciences Laboratory, University of California, Berkeley, CA 94720-7450, USA}

\author[0000-0002-5699-090X]{K. W. Paulson}
\affiliation{Smithsonian Astrophysical Observatory, Cambridge, MA 02138, USA}

\begin{abstract}
Recent in situ observations from Parker Solar Probe (PSP) near perihelia reveal ion beams, temperature anisotropies, and kinetic wave activity. These features are likely linked to solar wind heating and acceleration. During PSP Encounter 17 (at $11.4R_s$) on Sep-26-2023, the PSP/FIELDS instrument detected enhanced ion-scale wave activity associated with deviations from local thermodynamic equilibrium in ion velocity distribution functions (VDFs) observed by the PSP/Solar Probe Analyzers-Ion (SPAN-I). Dense beams (secondary populations) were present in the proton VDFs during this wave activity. Using bi-Maxwellian fits to the proton VDFs, we found that the density of the proton beam population increased during the wave activity and, unexpectedly, surpassed the core population at certain intervals. Interestingly, the wave power was reduced during the intervals when the beam population density exceeded the core density. The drift velocity of the beams decreases from 0.9 to 0.7 of the Alfv\'{e}n speed and the proton core shows a higher temperature anisotropy ($T_\perp/T_\parallel>2.5$) during these intervals. We conclude that the observations during these intervals are consistent with a reconnection event during a heliospheric current sheet crossing. During this event, $\alpha$ particle parameters (density, velocity, and temperature anisotropy) remained nearly constant. Using linear analysis, we examined how the proton beam drives instability or wave dissipation. Furthermore, We investigated the nonlinear evolution of ion kinetic instabilities using hybrid kinetic simulations. This study provides direct clues about energy transfer between particles and waves in the young solar wind.
\end{abstract}

\keywords{}

\section{Introduction} \label{sec:intro}
Kinetic-scale wave–particle interactions play a crucial role in the thermalization of the solar wind plasma \citep{Mar99,Kasper2002,XOV04,Hellinger2006,OV07,Klein2018}. The velocity distribution functions (VDFs) of the most-abundant plasma ions (protons and $\alpha$-particles) are often not in local thermodynamic equilibrium (LTE) and are subject to kinetic instabilities. Depending on $\beta_\parallel$, the instabilities could be driven by temperature anisotropy (R= \( T_\perp/T_\parallel \neq 1 \), where \( T_\perp \) and \( T_\parallel \) are the perpendicular and parallel temperatures with respect to the local magnetic field direction, respectively), super-Alfv\'{e}nic drifts between different particle populations, and temperature differences between different particle populations \citep{Gary1993}. These departures from LTE provide free energy to drive ion-scale waves. Analyzing particle VDFs and electromagnetic fields at scales comparable to the ion gyro radius reveals the sources of these wave–particle interactions. Recent studies using the Parker Solar Probe \citep[PSP;][]{Fox2016} data have explored the wave particle interactions using data, linear stability theory and hybrid modeling \citep[e.g.,][]{Bale2019, Bowen2020, Verniero2020, Verniero2022, Ofman2022, Ofman2023, McManus2024}. The conversion of magnetic and kinetic energy into plasma thermal energy at the ion scales occurs through wave-particle interactions such as ion-cyclotron resonances and other collisionless plasma instabilities. Collisions may play a significant role in solar wind thermalization on larger spatial and longer temporal scales, particularly far from the Sun at 1 AU \citep[e.g.,][]{Kasper2008, Kasper2017, Maruca2013, Johnson2024} where the effective proton mean free path is $\sim4\times10^5$ km or 0.6 $R_s$ \citep{Coburn2022}. However, close to the Sun in the young solar wind, collisions are negligible for the thermalization of non-Maxwellian plasma. Instead, a leading theory for the collisionless mechanisms of the thermalization process is through kinetic instabilities \citep{Kasper2008, Bourouaine2013, Martinovic2021}. The role of Coulomb collision, ion-cyclotron resonance heating, differential velocities of ion beams, and ion kinetic instabilities have been extensively studied in past research, particularly using Wind data near 1 AU \citep[215 $R_s$; e.g.,][]{Kasper2013, Kasper2017, Alterman2018} and Helios data \citep[e.g.,][]{Marsch1982a,Marsch1982b, Martinovic2021} as close as  0.3 AU ($\sim60 R_s$) to the Sun. These studies highlight the role of different kinetic processes in the heating solar wind plasma farther from the Sun compared to the present day PSP observations.

Plasma waves driven by the above-mentioned processes can be directly measured by spacecraft instruments using magnetic and electric field probes, provided that the temporal resolution is sufficient to capture the proton gyro-frequency range. These waves are often analyzed through magnetic field power spectra, magnetic field ellipticity (polarization), Poynting vectors, and other diagnostic methods \citep[e.g.,][]{He2011, Podesta2011a, Bowen2020a}. In the solar wind, three main types of ion kinetic instability-driven modes exist. Alfv\'{e}n/ion cyclotron (A/IC) instabilities occur due to temperature anisotropies in the core population of particles where $T_\perp / T_\parallel > 1$ and corresponding high value of $\beta_\parallel$, generating left-handed (LH) circularly polarized waves in the plasma rest frame. Fast magnetosonic/whistler (FM/W) instabilities, on the other hand, are driven by temperature anisotropies where $T_\perp / T_\parallel < 1$ and by super-Alfv\'{e}nic beams. These FM/W instabilities produce right-handed (RH) circularly polarized waves. In addition to the temperature anisotropies, the instability growth rates depend on the relative strength of the thermal pressure compared to the magnetic, parameterized as $\beta_\parallel$, as well as the relative abundance of the anisotropic species \citep[e.g., see][]{Gar93}. Additionally, ion/ion component instabilities arise from relative super-Alfv\'{e}nic drifts between different particle populations \cite[][and references therein]{McManus2024}. A comprehensive discussion of micro instabilities in the solar wind can be found in \cite{Verscharen2019}. 

Multi-ion plasma instabilities in the solar wind in the non-linear regime have been studied using hybrid particle-in-cell (hybrid-PIC) models, which treat ions as particles within a particle-in-cell framework while modeling electrons as a fluid \citep[e.g.,][]{XOV04,OV07,Ofman2010, Ofman2022,Ofman2023}. \citet{Maneva2013} employed a 1.5D hybrid model to demonstrate how protons and $\alpha$ particles are accelerated by large-amplitude Alfv\'{e}n wave spectra in the expanding solar wind plasma. Studies by \citet{Ofman2014, Ofman2017, Maneva2015, Ozak2015, Ofman2022, Ofman2023} use 2.5D hybrid models to investigate the effects of proton-$\alpha$ super-Alfv\'{e}nic drift in inhomogeneous expanding solar wind plasma and the evolution of turbulence at ion scales. More recently, several 3D hybrid models have been developed to study the turbulence cascade to ion scales in the solar wind plasma \citep{Vasquez2015, Cerri2017, Franci2018, Ofm19b,Hellinger2019, Markovskii2020, Vasquez2020,Markovskii2022}.

Although numerous studies have examined ion beams in the solar wind, usually, these studies assumed low-density beams relative to the core ion densities, consistent with previous observations. However, recent PSP observations at E17, as reported in this paper, have identified streams with comparable beam-core population densities, accompanied by ion-scale waves, thereby motivating the present study. Since these beams are likely associated with  (or 'leaked from') the reconnection events in heliospheric current sheet, these observations can provide valuable insights into how reconnection events transfer energy to the solar wind plasma on ion-scales. Furthermore, this study explores the understanding of the energy exchange mechanisms between waves and particles. The paper is structured as follows: Section \ref{sec:data} presents the data and model; Section \ref{sec:obs} highlights the observations of dense beams, including their temperature anisotropy and density calculations; Section \ref{sec:ls} describes the linear stability analysis; and Section \ref{sec:mod} focuses on modeling these dense beams with nonlinear simulations. Finally, discussions and conclusions are provided in Sections \ref{sec:dis}.

\section{Dataset and Model}\label{sec:data}
\subsection{Data}
Non-Maxwellian features, such as beams and temperature anisotropies, have been detected in the solar wind at perihelia by SPAN-I, part of the SWEAP instrument suite on PSP \citep{Kasper2016, Livi2022}. SWEAP aims to study the sources of the solar wind, heating processes, and particle transport. It includes SPAN-I, two electron analyzers \citep[SPAN-E;][]{Whittlesey2020}, and a Sun-pointing Faraday cup, Solar Probe Cup \citep[SPC;][]{Case2020}. Collectively, these instruments provide nearly complete sky coverage during most of an encounter, with only small gaps in ion detection \citep{Kasper2016, Kasper2019}.

This study focuses on data from SPAN-I, a top-hat electrostatic analyzer and mass discriminator, to measure 3D ion VDFs within an energy range of 2 eV to 30 keV. SPAN-I’s field of view (FOV) spans $247.5^\circ \times 120^\circ$ and utilizes a time-of-flight section to distinguish protons, $\alpha$-particles, and heavy ions. However, part of its FOV is blocked by the PSP’s thermal shield, resulting in ``partial moments.'' SPAN-I is particularly useful for identifying ion beams because nonthermal features often appear at higher velocities, which are more likely to fall within its FOV. For the time interval analyzed in this paper, the proton core peak remains mostly within the instrument's FOV for nearly the entire duration. However, the observations of $\alpha$ particle VDFs are limited by the partial FOV \citep{Mostafavi_2022, Mostafavi2024}. Therefore, we have used $\alpha$ particle moment data to show the general variations in $\alpha$ particle properties, and do not include  $\alpha$ particle VDFs  in our detailed analysis.

Electric and magnetic fields are measured by the FIELDS instrument onboard PSP. The FIELDS instruments include electric antennas and fluxgate magnetometers \citep[MAG;][]{Bale2016}. The MAG data have an observational cadence of approximately $\sim$293 Hz (near perihelion), more than enough to resolve the proton gyroresonance frequency at perihelion ($\sim 10$Hz). The elctromagnetic wave properties of the solar wind plasma are derived using the MAG data.

\subsection{Hybrid Model}
\label{hybmodel:int}
We have used the well-tested parallelized 2.5D hybrid-PIC (hereafter referred to as the hybrid model) code to model multi-ion, magnetized solar wind plasma. The hybrid model computes the self-consistent evolution of the ion VDFs, including nonlinear wave-particle interactions, and ion kinetic instabilities in magnetized near-Sun solar wind and other analogous plasma systems. The present code builds on the 1D hybrid model by \cite{WO93}, later extended to 2D by \cite{Ofman2007}, parallelized by \cite{Ofman2010} and extended to 3D by \cite{Ofman2019}. The hybrid modeling approach captures ion dynamics across many ion-cyclotron periods with real proton mass and Alfv\'{e}n speed to light speed ratio (an advantage compared to common approximations in PIC codes), while representing electrons as massless neutralizing fluid.  The electric field is obtained from a generalized Ohms’ law, and the magnetic field from the solution of Maxwell’s equations using the psedo-spectral method with periodic boundary conditions. The particle and field equations are  integrated in time using the Rational Runge-Kutta method \citep{wambecq1978}. Excluding the electron inertia provides computational efficiency but limits the frequency range to around proton cyclotron frequency. In the present study, we use the 2.5D hybrid code with three velocity, magnetic, and electric field components in two spatial dimensions. This allows modeling obliquely-propagating as well as parallel-propagating wave modes. Similar to recent studies \citep{Ofman2022, Ofman2023, Ofman2025}, we have applied a $256^2$ spatial grid with up to 256 particles per cell, effectively capturing plasma dispersion and evolution in the ion scale range. Each numerical particle represents a large number of real particles, determined by the density normalization. The number of particles per cell is chosen to control statistical noise and ensure proper VDF velocity-space resolution, which typically depends on plasma \( \beta \). High-order filtering terms are applied as needed to minimize numerical noise. Numerical stability and energy conservation are verified through convergence tests, ensuring noise remains below physical fluctuation levels in the hybrid modeling results. This numerical method has been validated in multiple solar wind modeling studies \citep[e.g.,][]{Ofman2010, Ofman2022, Ofman2023}.  The results obtained from the model are discussed in Section~\ref{sec:mod}.

\section{Observation of dense beams}\label{sec:obs}

\begin{figure}[h]
\centering
(a)\\
\includegraphics[width=0.8\linewidth]{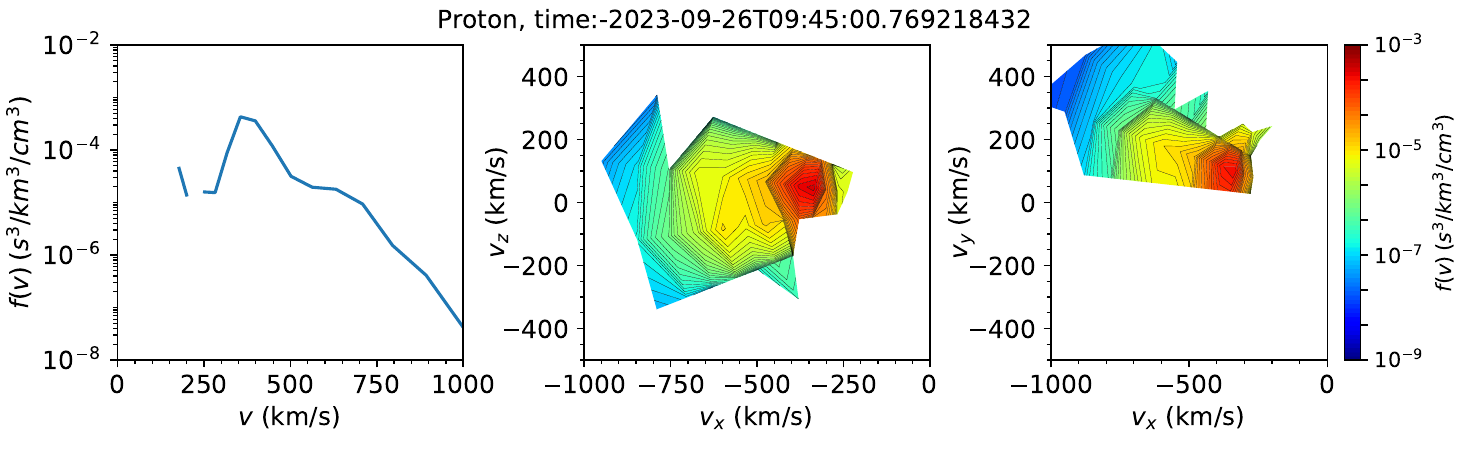}\\
(b)\\
\includegraphics[width=0.8\linewidth]{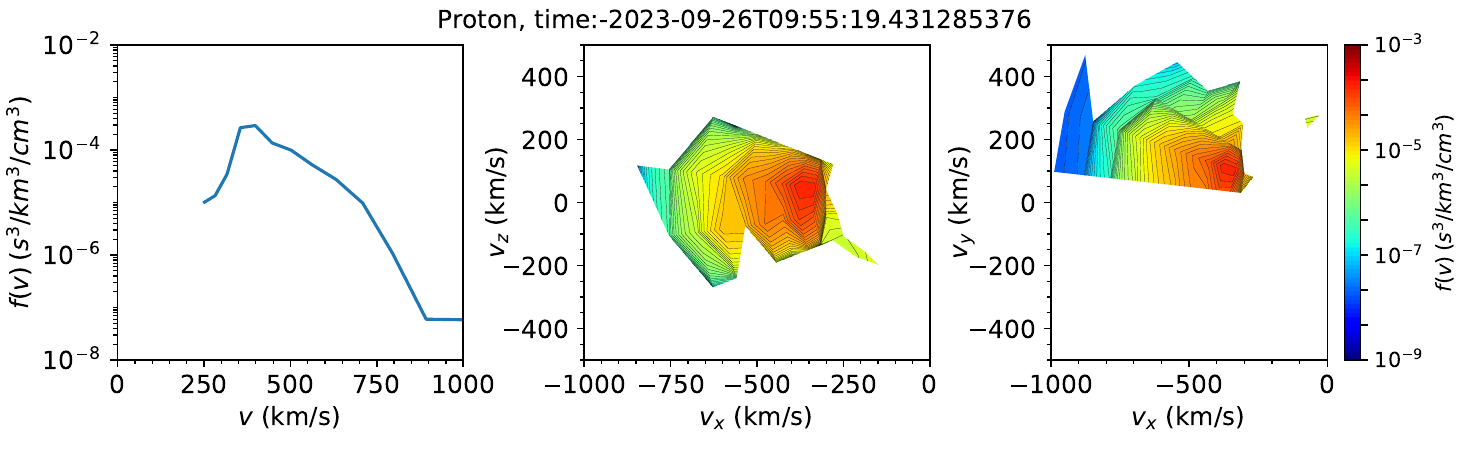}\\
(c)\\
\includegraphics[width=0.8\linewidth]{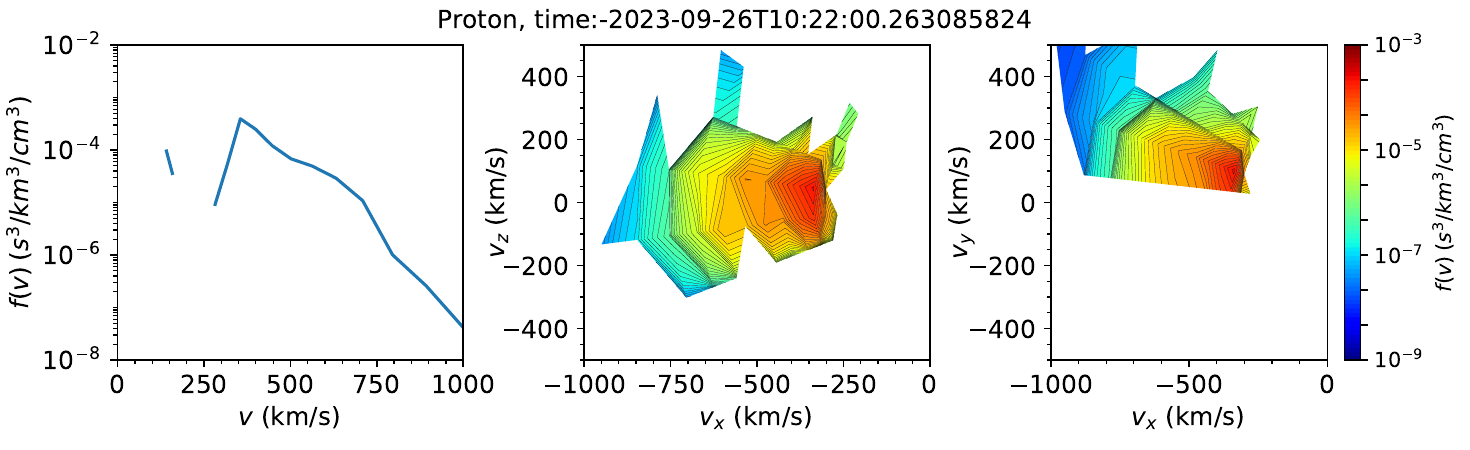}\\
(d)\\
\includegraphics[width=0.8\linewidth]{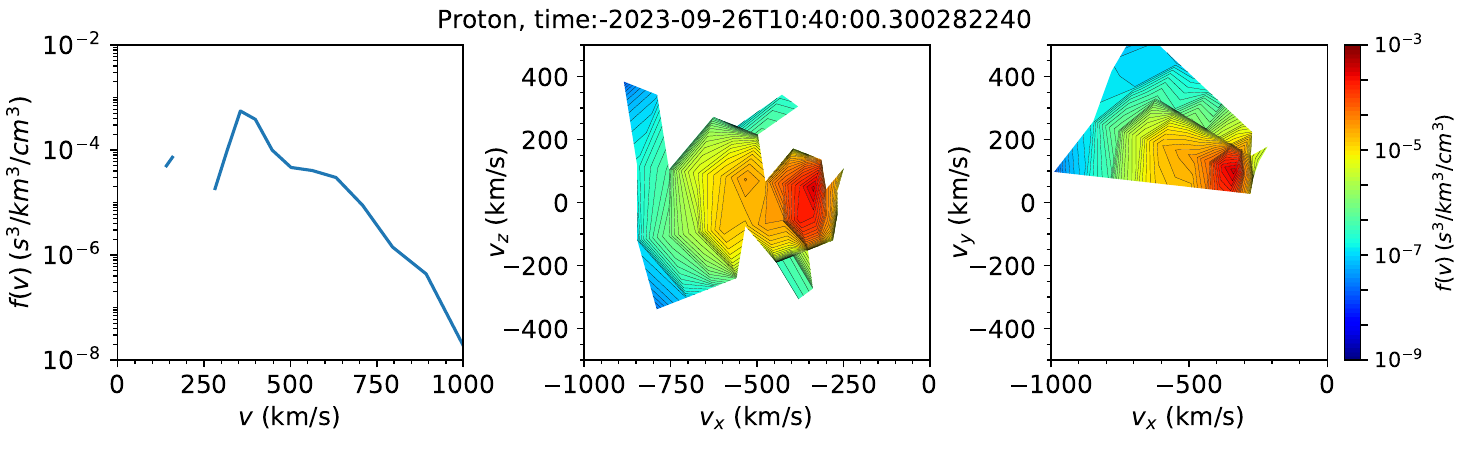}
\caption{First column of panels (a)-(d) shows the VDF summed over all look directions, the second column shows the VDF contours summed and collapsed onto the $\theta$  plane, the third column shows the VDF contours summed and collapsed onto the azimuthal plane. Panels (b) and (c) have a beam-to-core density ratio of greater than unity. The discontinuities in the first column are observational artifacts caused by data gaps observed in the detector.}

\label{fig:vdf}
\end{figure}

This section presents observations of dense beams using PSP/SPAN-I data at Encounter 17. A 3D bi-Maxwellian function fit to the SPAN-I data has been used to determine the density and temperatures ($T_\parallel,\ T_\perp$) of protons, following the methodology outlined in \cite{McManus2022, McManus2024}. In this study, we adopt the convention that the population with the higher phase space density, given by the bi-Maxwellian normalization constant
\begin{equation}\label{eq:1}
f = n \left( \frac{m}{2 \pi} \right)^{\frac{3}{2}} \frac{1}{T_{\perp} T_{\parallel}^{1/2}}    
\end{equation}
with the peak near the solar wind centroid moving frame at its bulk velocity is labeled as the core. In Equation \ref{eq:1}, $n$ denotes the number density, $m$ is the mass, and $T_\perp$ and $T_\parallel$ represent the perpendicular and parallel temperatures, respectively. The observed VDFs (see Figure \ref{fig:vdf}) clearly show that while the beam has a lower phase space density than the core, it exhibits a significantly larger \(T_\parallel\), which can lead to an increased beam ion number density.

\begin{figure}
\begin{center}
\plotone{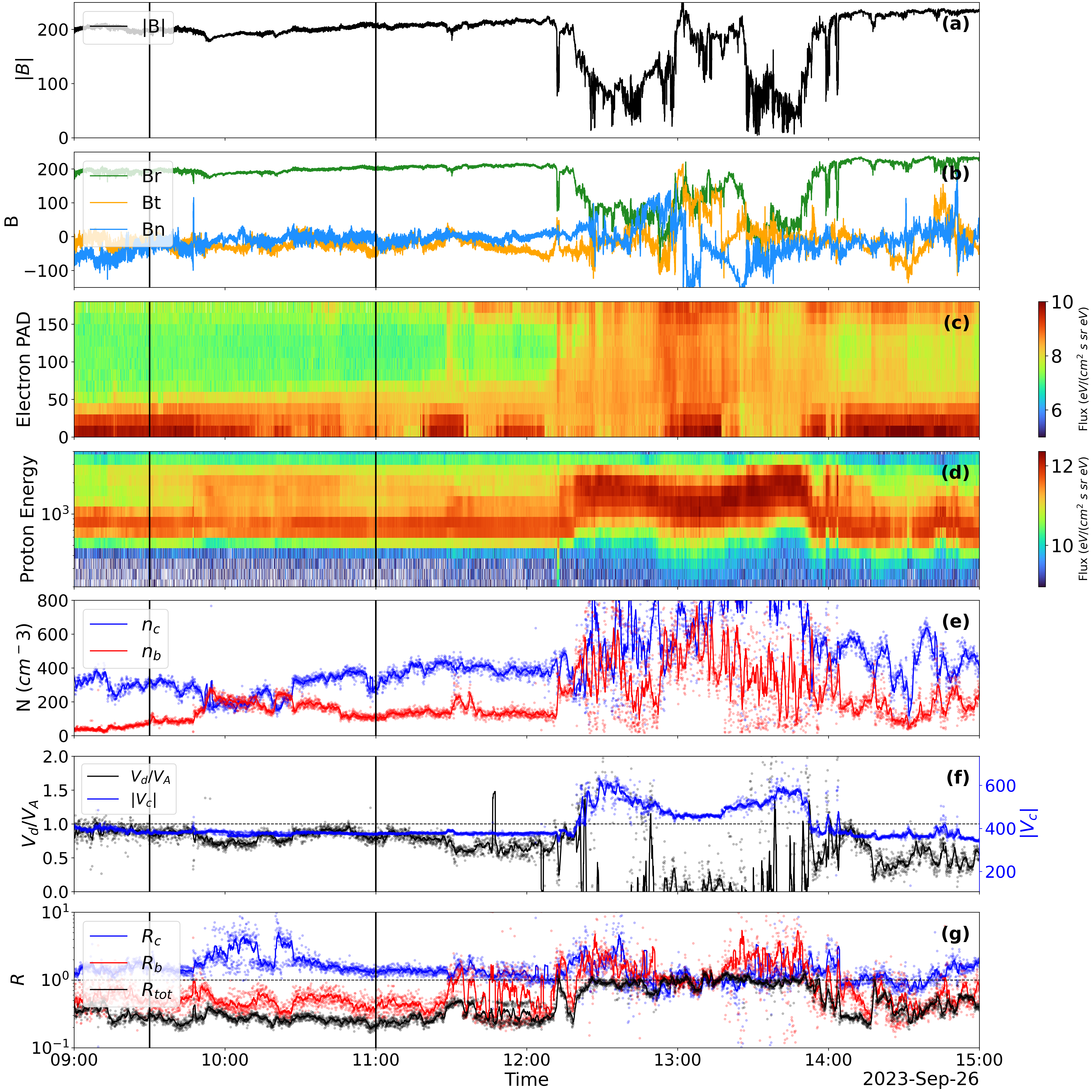}
\caption{\small From top to bottom, the variation of the magnetic field magnitude, the magnetic field components ($B_r$-green, $B_t$-yellow, $B_n$-blue), electron pitch angle distribution (PAD) at 432 eV, proton flux variation with energy, proton core number density (blue) and proton beam number density (red), the ratio of the differential velocity between proton core and beams to the Alfv\'{e}n velocity (black) and proton core velocity (blue), temperature anisotropy of proton core (blue), beam (red) and total (black) are shown in panel (a) to (g). The vertical black lines show the time period of observed relatively dense beams. These observations are from PSP encounter 17.   \label{fig:cs}}
\end{center}
\end{figure}

\begin{figure}
\begin{center}
\plotone{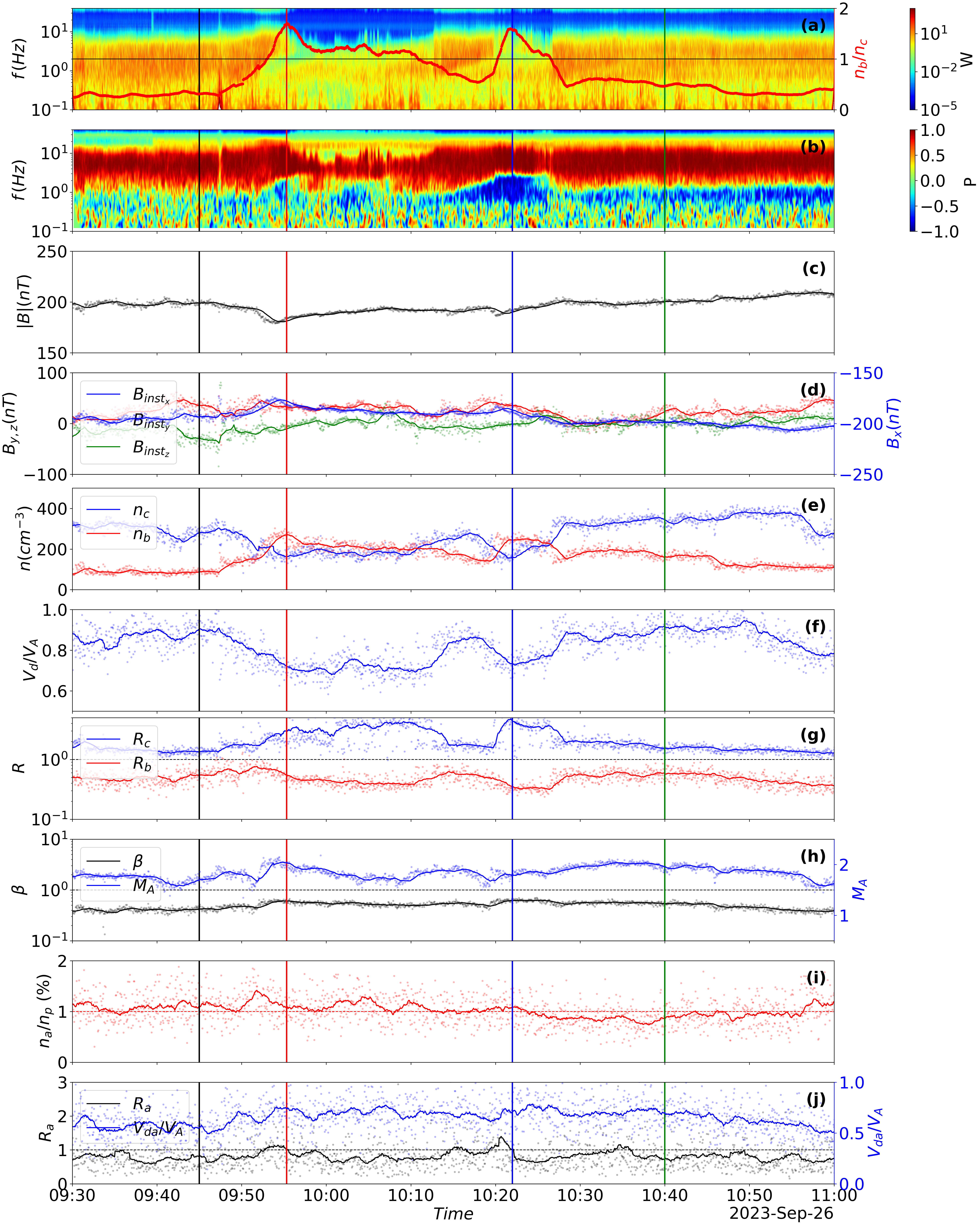}
\caption{\small From top to bottom, the panels (a)-(j) show the variation of wavelet power spectra (W), wave polarization (P), magnetic field magnitude, magnetic field component in instrument frame ($B_x$ in blue, $B_y$ in red, and $B_z$ in green), proton core ($n_c$ - blue) and beam ($n_b$ - red) number densities, the drift velocity of proton beams normalized to the Alfv\'{e}n velocity ($V_d/V_A$), temperature anisotropies of the proton core ($R_c$ - blue) and beam ($R_b$ - red), plasma beta ($\beta$ - black) and Alfv\'{e}n Mach number ($M_A$ - blue), the $\alpha$-to-proton number density ratio ($n_a/n_p$), the temperature anisotropy of $\alpha$ particles ($R_a$ - black), and the drift velocity of $\alpha$ particles relative to the Alfv\'{e}n velocity ($V_{da}/V_A$- blue)  are shown. The vertical black, red, blue, and green lines show the time period for which model outputs are discussed in section~\ref{sec:mod}.     \label{fig:eve}}

\end{center}
\end{figure}

Figure~\ref{fig:cs} presents the PSP observation of a part of current sheet around 1230 UT to 1400 UT (full current sheet is not shown) at Encounter 17 on 26 September 2023.  The panels in Figure \ref{fig:cs}, show the magnetic field magnitude, the magnetic field components ($B_r$ - green, $B_t$ - yellow, $B_n$ - blue), electron strahl Pitch Angle Distribution (PAD at 433eV), proton flux variation with energy, proton core number density (blue) and proton beam number density (red), the ratio of the differential velocity between the proton cores and beams to the Alfv\'{e}n velocity (black) and core velocity (blue), and the temperature anisotropy of proton cores (blue), beams (red), and all the protons (black) calculated using core and beam temperature. The solid lines in each panel represents the 15 data points moving mean. The panel (d) shows the proton flux associated with (or 'leaked' from)  the current sheet around 09:00\,UT before the start (around 12:30\,UT) of the current sheet reconnection event. This can be inferred from the fluxes at different energy levels. The energy of the leaked protons is similar to that of the protons observed in the current sheet. A similar observation of leaked protons from PSP Encounter 8 is studied in \cite{Phan2022}. Panel (e) shows the beam density exceeding the core density for some time between 09:30\,UT and 11:00\,UT. This region of interest is further explored in Figure~\ref{fig:eve}.

Figure \ref{fig:eve} illustrates the observations of wave and particle properties. The methodology described in \cite{Shankarappa2023} is followed to calculate the wave properties. The Morlet wavelet is used to calculate the wave parameters. Panel (a) shows the wavelet Power Spectra (W) of the waves, where the red curve indicates the beam-to-core density ratio on the right $y$-axis. Panel (b) depicts the polarization (P) of the waves, highlighting the presence of left-hand circularly polarized waves (in red), which exhibit variations in the frequency band as the beam-to-core density ratio changes. Additionally, right-hand waves (in blue) are also observed in certain intervals (after 1010 UT).

Panels (c) to (j) present the magnitude of the magnetic field ($|B|$), its components in the instrument frame ($B_x$ in blue, $B_y$ in yellow, and $B_z$ in green), proton core and beam number densities ($n_c$ in blue and $n_b$ in red, respectively), drift velocity of proton beams normalized to the Alfv\'{e}n velocity ($V_d/V_A$), temperature anisotropies of the proton core and beam ($R_c$ in blue and $R_b$ in red, respectively), plasma beta ($\beta$ in black) and Alfv\'{e}n Mach number ($M_A$ in blue), the $\alpha$-to-proton number density ratio ($n_a/n_p$, where $n_p=n_c+n_b$), the temperature anisotropy of $\alpha$ particles ($R_a$ in black), and the drift velocity of $\alpha$ particles relative to proton core normalized by the Alfv\'{e}n velocity ($V_{da}/V_A$ in blue).

From Figure~\ref{fig:eve}, it is evident that during intervals where the beam number density exceeds the core number density: (i) the wave frequency band narrows, (ii) the magnetic field magnitude slightly decreases, (iii) the drift velocity between the core and beam decreases, (iv) the temperature anisotropy of the core increases, and (v) the beam temperature anisotropy decreases. The plasma $\beta$ remains below 1 throughout the observed period. The observed region contains super-Alfv\'{e}nic solar wind, as the Alfv\'{e}n Mach number exceeds 1.

\begin{table}[ht]
\caption{The dimensionless parameters of the numerical hybrid model runs for the various cases in the present study. The initial beam-core drift speed  $V_{dp}$ for protons, the initial proton core density $n_{pc}$, the initial proton beam density $n_{pb}$, the $\beta_{||}$ of the ion populations (defined in terms of $n_e$), initial temperature anisotropy of proton core ($R_{pc}$) and beams ($R_{pb}$), beam to core number density ratio ($bc$) and Alfv\'{e}nic velocity ($v_A = \frac{B}{\sqrt{\mu_0 \rho}}$) is shown, respectively. The last column shows the times of the PSP observations corresponding to the selected cases.}
\centering
\begin{tabular}{|l|c|c|c|c|c|c|c|c|c|c|}
\hline
Case \# & $V_{dp}$  & $n_{pc}$  & $n_{pb}$ & $\beta_{pc}$ & $\beta_{pb}$ & $R_{pc}$ &  $R_{pb}$ & $bc$ &$v_A (km/s)$& Time\\
\hline
1 (Black) &  0.961  &  0.765 & 0.235  & 0.060 & 0.398 & 1.349 & 0.503 & 0.31 &219.4 & 2023-09-26 09:45:00.8\\
2 (Red) &  0.668  &  0.362 & 0.638  &  0.044 &0.540 &  2.26 & 0.420 & 1.76 &196.5 & 2023-09-26 09:55:19.4\\
3 (Blue) &  0.679  &  0.410 & 0.590  &  0.047 & 0.510 &  1.407 & 0.380 & 1.44 &209.6 & 2023-09-26 10:22:00.3\\
4 (Green) &  0.945  &  0.661 & 0.339    &  0.059 & 0.447 & 1.406 & 0.424 & 0.51 & 198.2 & 2023-09-26 10:40:00.3\\
\hline
\end{tabular}

\label{model_param:tab}
\end{table}

The $\alpha$ particle parameters calculated from the moments are shown in panels (i) and (j). The $\alpha$-to-proton density ratio, $\alpha$ particle temperature anisotropy, and the $\alpha$ particle drift velocity relative to the proton core (in units of $V_A$) remain nearly constant throughout the interval.
  
To investigate the relationship between these strong beams and the observed waves, we performed linear stability calculations, as discussed in Section \ref{sec:ls}. Additionally, to examine the evolution of temperature anisotropy and energy exchange between waves and particles, we employed a nonlinear hybrid model. The vertical-colored lines in Figure~\ref{fig:eve} mark the times used as input for the hybrid model, with the input details provided in Table \ref{model_param:tab}. The proton VDFs associated with these intervals are shown in Figure~\ref{fig:vdf}. The observed VDFs provide significant insight into the behavior of dense beam cases. In all instances, the VDF peak is highest at the core, offering a clear definition of what constitutes the core VDF. Additionally, the beam VDF is consistently broader (hotter) than the core VDF in the parallel direction ($T_\parallel$). This is further supported by the observation that $R_c > 1$ while $R_b < 1$ for all times shown in Figure \ref{fig:eve}. Panels (b) and (c) show that the beam distributions become elongated along the parallel direction, causing the beam density (i.e., the integrated beam VDF) to exceed the core density (i.e., the integrated core VDF). This apparent parallel heating of the beam is associated with a cooler core, leading to higher $R_c$ and lower $R_b$ in these regions compared to regions with smaller beam density. 


While the comparable densities of the beam and core could arise from different process, including isotropisation, here, in fact it is likely a signatures of non-equilibrium, as evident from the VDF structure. After inspecting the proton VDF derived from SPAN-I data in Figure \ref{fig:vdf}, it is clear that the structures of the core and beam VDFs are very different. The core predominantly exhibits a bi-Maxwellian structure with $T_{\perp}/T_{\parallel} > 1$, featuring a clear peak near the low end of the velocity space at approximately 350 km/s. In contrast, the beam displays a significantly larger $T_{\parallel}$ with a ``plateau" in velocity space between 500--700 km/s, where $T_{\parallel,b} \gg T_{\parallel,c}$, and the beam VDF extending to higher velocities $\sim$1000 km/s. As a result, the lower $T_{\parallel,c}$ produces a more narrow distribution of the core, which contains a comparable number of protons to the much broader distribution of the beam. Since the core and beam densities are calculated by integrating the VDFs over velocity space, this leads to the  outcome that the beam and core exhibit similar densities in this observation.


\section{Linear Stability Analysis}\label{sec:ls}

Using linear stability analysis, we investigate which plasma waves could account for the observed left-handed polarization in Figure \ref{fig:eve} and determine the associated dissipation and heating rate using the \texttt{PLUME} (Plasma in a Linear Uniform Magnetized Environment) linear plasma dispersion solver \citep{Klein_Howes_2015_PLUME}.
\texttt{PLUME} calculates the hot linear plasma dispersion relation $|D(\omega, \gamma; \mathbf{k}, \mathcal{P})| = 0$ for relatively drifting bi-Maxwellian VDFs of ions and electrons in a uniform magnetized environment. 
Here $\mathcal{P}$ is the underlying set of equilibrium plasma parameters that describe the constituent VDFs. 
The solutions of the dispersion relation provide dimensionless angular frequencies and damping rates, $(\omega/\Omega_p, \gamma/\Omega_p)_{\texttt{PLUME}}$ in the plasma frame, along with the corresponding eigenfunctions and damping rates associated each ion and electron population. 
Here $\Omega_p$ denotes the proton gyrofrequency. 
Since the $\alpha$ particle density remains low compared to proton density throughout the interval, we focus on the proton contributions and consider the parameter space $\mathcal{P} = \left[ \beta_{\parallel,c}, \frac{T_{\parallel,c}}{T_{\parallel,b}}, \frac{T_{\perp,c}}{T_{\parallel,c}}, \frac{T_{\perp,b}}{T_{\parallel,b}},  \frac{V_d}{V_A}, \frac{T_{\parallel,c}}{T_{\parallel,e}} \right]$.  
The electrons are assumed to be isotropic and also drifted sufficiently to achieve zero net current, with $T_e=T_{\parallel,c}$. 

\begin{figure}
    \centering
    \includegraphics[width=0.9\linewidth]{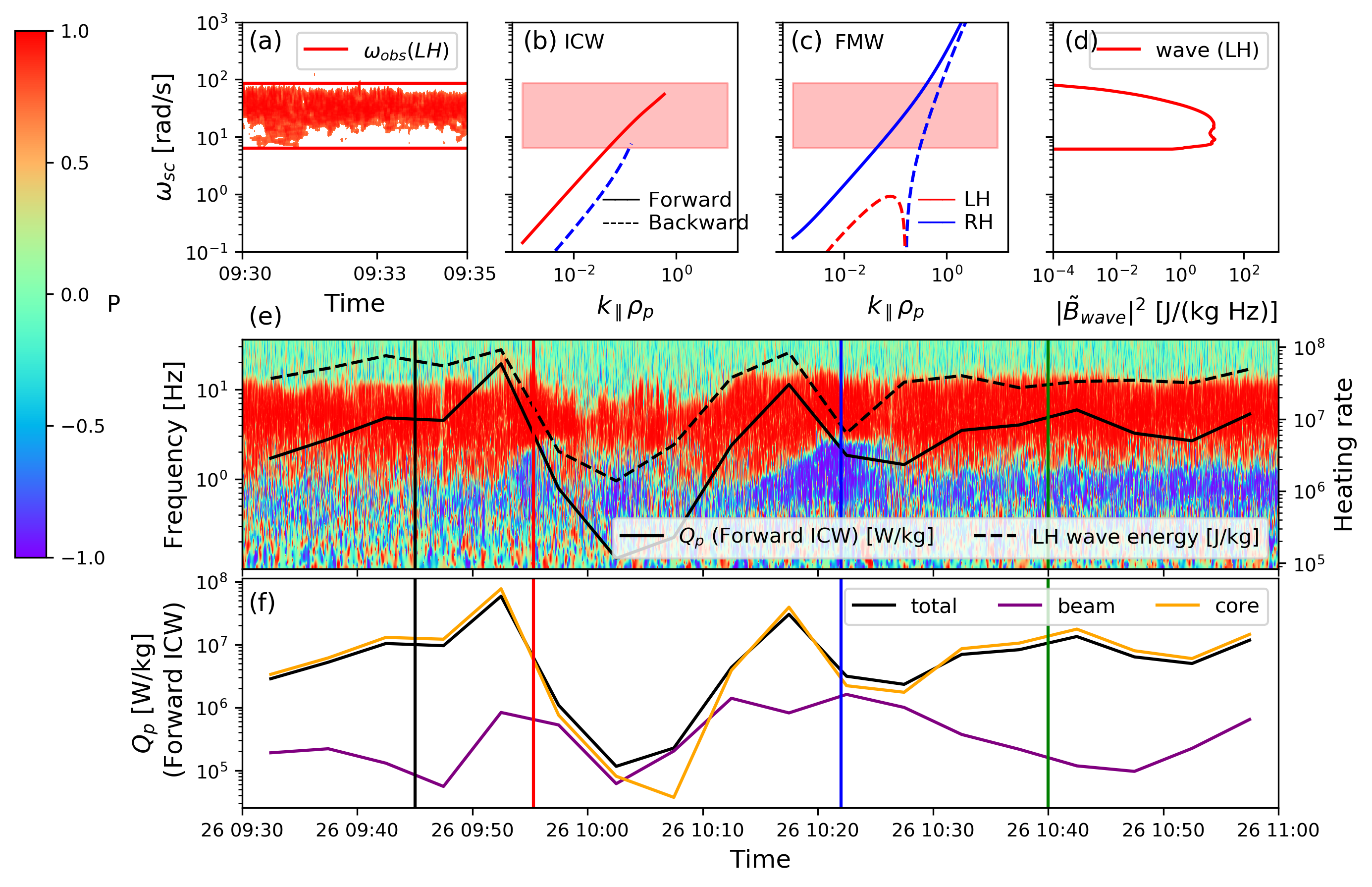}
    \caption{Illustration of the routine to estimate the dissipation rate of the observed LH wave considered in Figure \ref{fig:eve} using  \texttt{PLUME} linear plasma dispersion solution for a typical 5-minute interval: (a) Identified band of frequencies where LH circular polarization is observed. (b,c) The Doppler shifted frequencies of sunward and anti-sunward ICWs (b) and FMWs (c) as a function of parallel wavevector. (d) The energy spectrum of the observed LH wave. (e) The time profile of the estimated heating rate, $Q_p$ (black solid), and the LH wave energy (black dashed) as well as (f) the partitioning of $Q_p$ into proton core (orange) and beam (purple).}
    \label{fig:linear analysis}
\end{figure}

We subdivide the interval considered in Figure \ref{fig:eve} into smaller 5-minute intervals. 
For each smaller interval, we identify $\omega_{obs}$, the angular frequency band where circularly circularly polarized waves occur (red shaded region in Panel (a) of Figure \ref{fig:linear analysis}). We then distinguish their energy, $|\tilde{B}_{\textrm{wave}}(\omega_{sc})|^2$, from the background turbulence using polarization, $P$, as a criterion (Panel (d)) as described in \cite{Shankarappa2023} and \cite{Shankarappa2024}. Here, $\omega_{sc}$ are angular frequencies in the spacecraft frame.
Both parallel propagating Ion cyclotron waves (ICWs, LH in plasma frame, c.f. see Appendix C of \cite{Shankarappa2024} for a discussion of polarization and handedness of observed waves) and fast magnetosonic waves (FMWs, RH in plasma frame) can appear as either RH or LH in spacecraft frame due to the Doppler shift, making them difficult to distinguish. 
We compute the plasma frame frequencies and damping rates of forward and backward propagating ICWs and FMWs for local values of $\mathcal{P}$.
We then determine the Doppler shifted frequencies, considering all possible shifts due to both sunward and anti-sunward propagating ICWs and FMWs (as described in \cite{Shankarappa2024}). The results are shown in Panels (b) and (c) for a typical 5-minute sub-interval that is representative of the whole interval considered in Figure \ref{fig:eve}. 
Only the forward ICW Doppler-shifted frequencies overlap with the frequency band at which the LH wave is observed and have the correct polarization (red curve and shaded region in Panel (b)). 
Next, we estimate the dissipation rate of the observed waves onto protons, $Q_p = \int_{\omega_{obs}} |\tilde{B}_{\textrm{wave}}(\omega_{sc})|^2 \gamma_p(\omega_{sc})$, where $\gamma_p(\omega_{sc})$ is the linear damping rate onto the protons shifted into the spacecraft frame frequencies that overlap with $\omega_{obs}$. The terms 'heating rate' and 'dissipation rate' are used interchangeably to refer to the same quantity, $Q_p$. The $Q_p$ values are higher when the wave energy is higher (as seen in the correlation of solid and dashed curves in panel (e)) and/or the peak of the damping rate profile matches well with the peak of the wave energy spectrum.

By repeating the estimation of $Q_p$ for all 5-minute intervals, we evaluate its time evolution (black curve in Panels (e) and (f)). 
We observe that the sharp decrease in $Q_p$ is correlated by a steep drop in wave energy (black dashed in Panel (e)) and polarization that corresponds to increased beam density. 
We further estimate the partitioning of $Q_p$ onto the proton beam and core (purple and orange, respectively, in Panel (f)).
In the intervals with relatively diffuse beams, we infer that core protons experience dominant heating.
When the beam population is denser than the core, power is more evenly partitioned between the two populations. However, the overall dissipation rate decreases due to the lower wave power.
By coupling the linear dispersion solution determined from local plasma conditions with the observed wave power, we are able to estimate dissipation rates and infer a significant relative contribution in heating due to a denser proton beam. This contribution is closely associated with the significant reduction of the wave power. 
We next conduct a nonlinear evaluation of such solar wind plasma to investigate their evolution.

\section{Hybrid Modeling Results}\label{sec:mod}

In this study, we employ the hybrid model to simulate two ion species: proton cores and proton beams using the methodology described in Section~\ref{hybmodel:int}. The hybrid model computes the self-consistent evolution of ion velocity distribution functions (VDFs), capturing wave-particle interactions and ion kinetic instabilities. It builds on earlier versions by \citet{WO93}, \citet{Ofman2007}, \citet{Ofman2010}, and \citet{Ofman2019}, and treats ions as particles by solving the ion equations of motions subject to the Lorentz force of large number ($10^7-10^8$) of numerical particles and electrons as a massless neutralizing fluid. The electric field is derived from a generalized Ohm’s law, and the magnetic field is evolved using a pseudo-spectral method with periodic boundaries. The particle and field equations are integrated using the Rational Runge-Kutta method. We use a $256^2$ grid with up to 256 particles per cell for each ion species, sufficient to capture ion-scale dynamics and reduce numerical noise. This 2.5D setup includes three components of velocity, magnetic, and electric fields in two spatial dimensions, allowing both parallel and oblique wave propagation with background magnetic field direction (defined as $B_0$ in the x-direction). High-order filtering and convergence tests ensure numerical stability, and the model has been validated in previous solar wind studies. The normalized input parameters used in the hybrid model are given in Table~\ref{model_param:tab}. In the present hybrid model, we define two proton populations and track the evolution of each corresponding VDF. In the initial state the particles are labeled as belonging to the core or beam population. Throughout the simulation, it is known whether particles belong to the core or the beam population and the evolution of the beam and core particles is tracked. At each stage of the evolution, the VDF is fitted with a bi-Maxwellian function to calculate the relevant plasma parameters e.g., thermal velocity, drift velocity, density etc. The thermal velocity is used to get the thermal energy which indicate if VDF is widening or shrinking. The drift velocity is an indicator of the nonthermal kinetic energy which indicate how the VDFs evolve in the parallel (to the magnetic field) velocity space. These examples are chosen based on beam to core density ratio in the PSP/SPAN-I data. We classify the four cases into two groups of modeling parameters: Class~I, which includes Case 1 (black) and Case 4 (green), where the core number density is higher than the beam number density; and Class~II models, which includes Case 2 (red) and Case 3 (blue), where the beam number density exceeds the core number density. Additionally, Class~I corresponds to intervals with a larger frequency band or more waves, whereas Class~II models represents intervals with a smaller frequency band or fewer waves, as illustrated in Figure \ref{fig:eve}.  

\begin{figure}[h]
\centering
(a)\hspace{2in}(b)\hspace{2in}(c)\\
\includegraphics[width=0.32\linewidth]{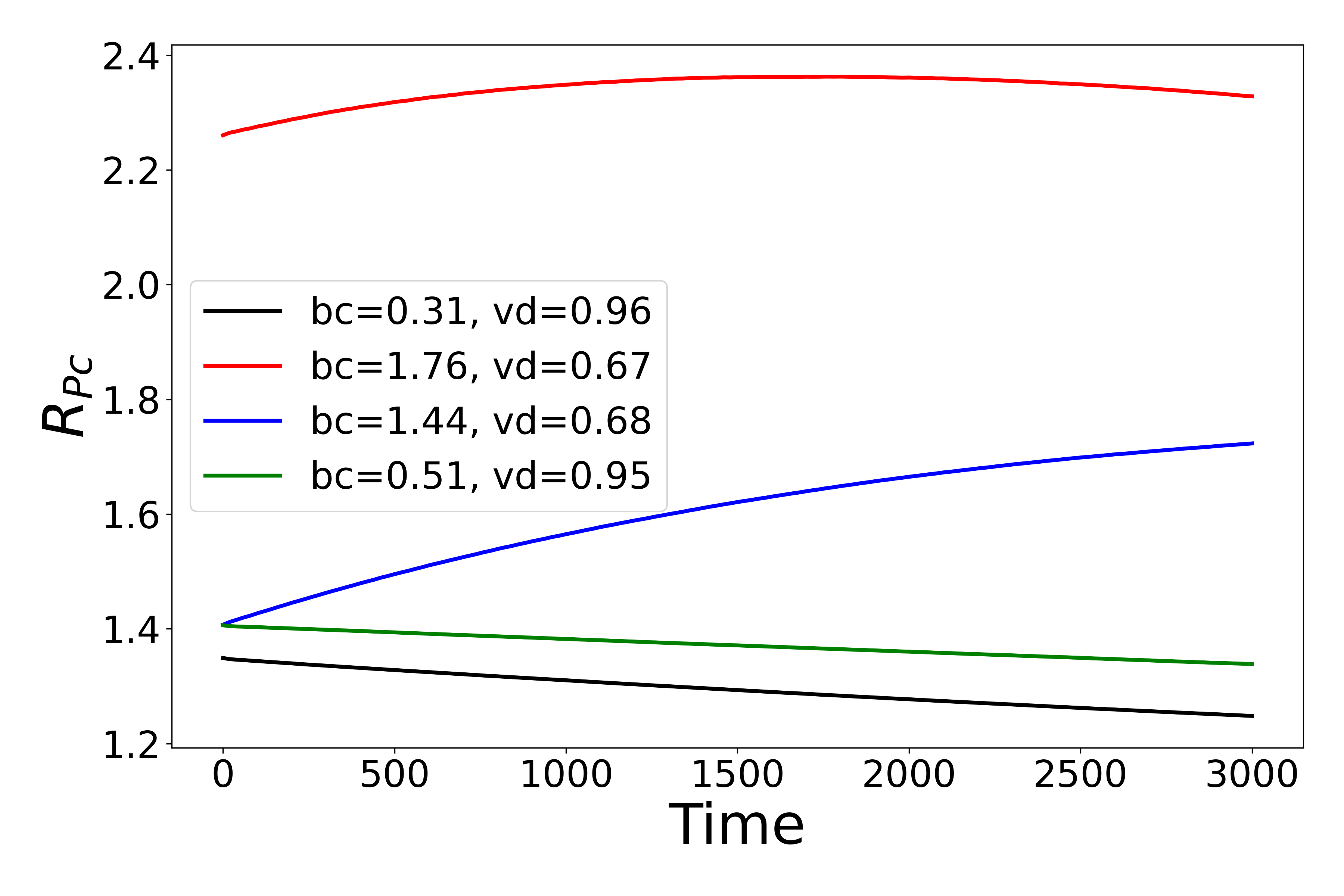}
\includegraphics[width=0.32\linewidth]{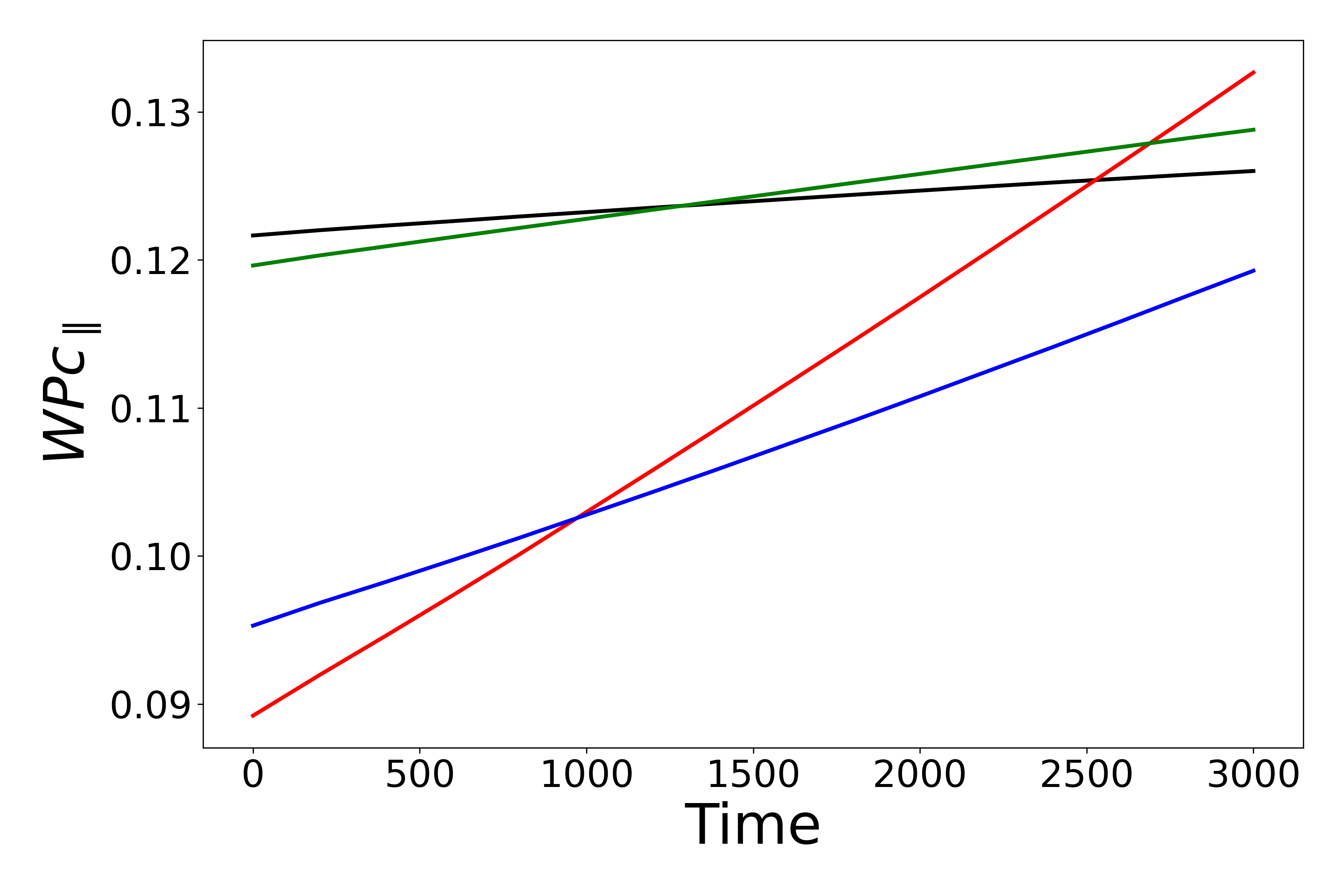}
\includegraphics[width=0.32\linewidth]{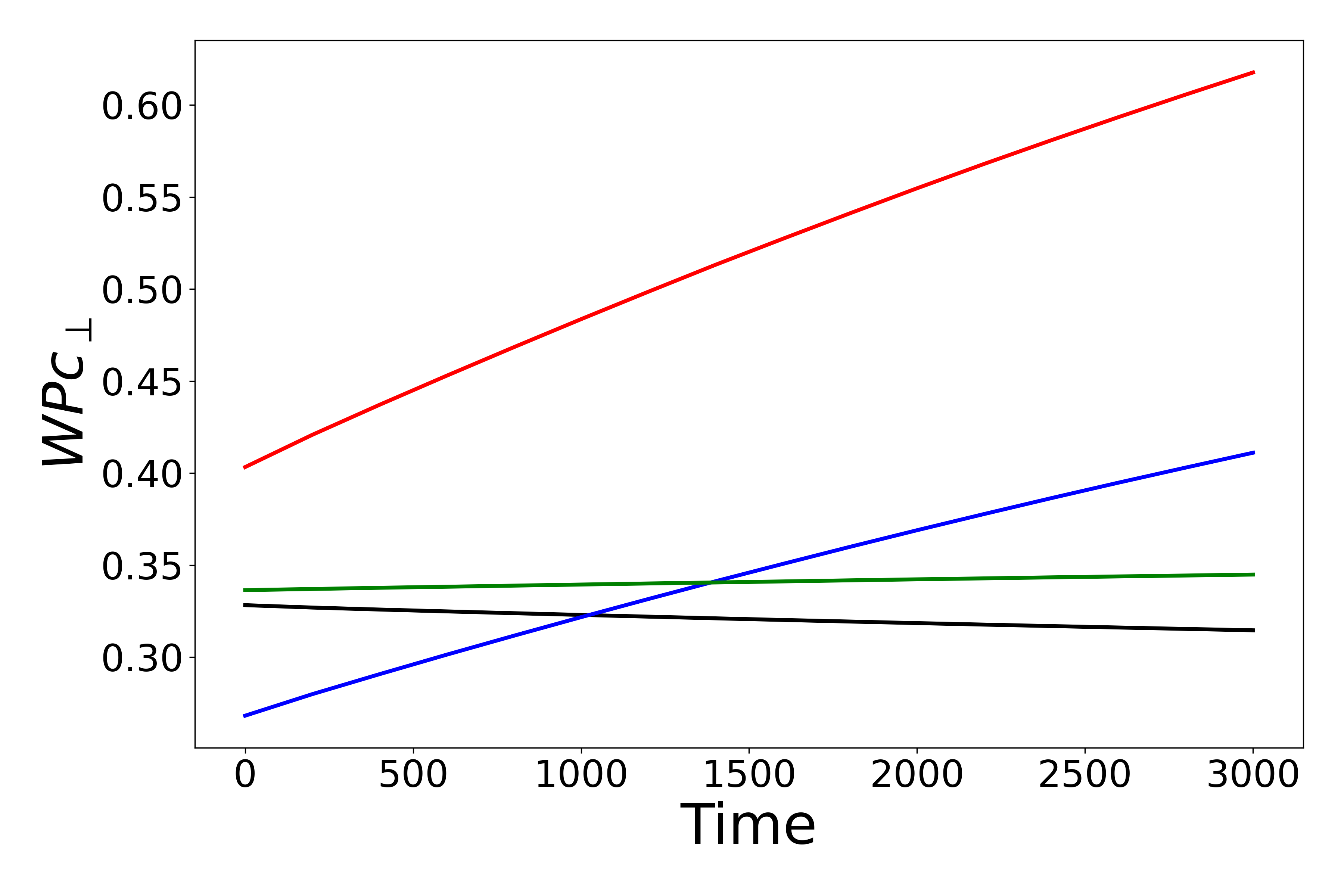}\\
(d)\hspace{2in}(e)\hspace{2in}(f)\\
\includegraphics[width=0.32\linewidth]{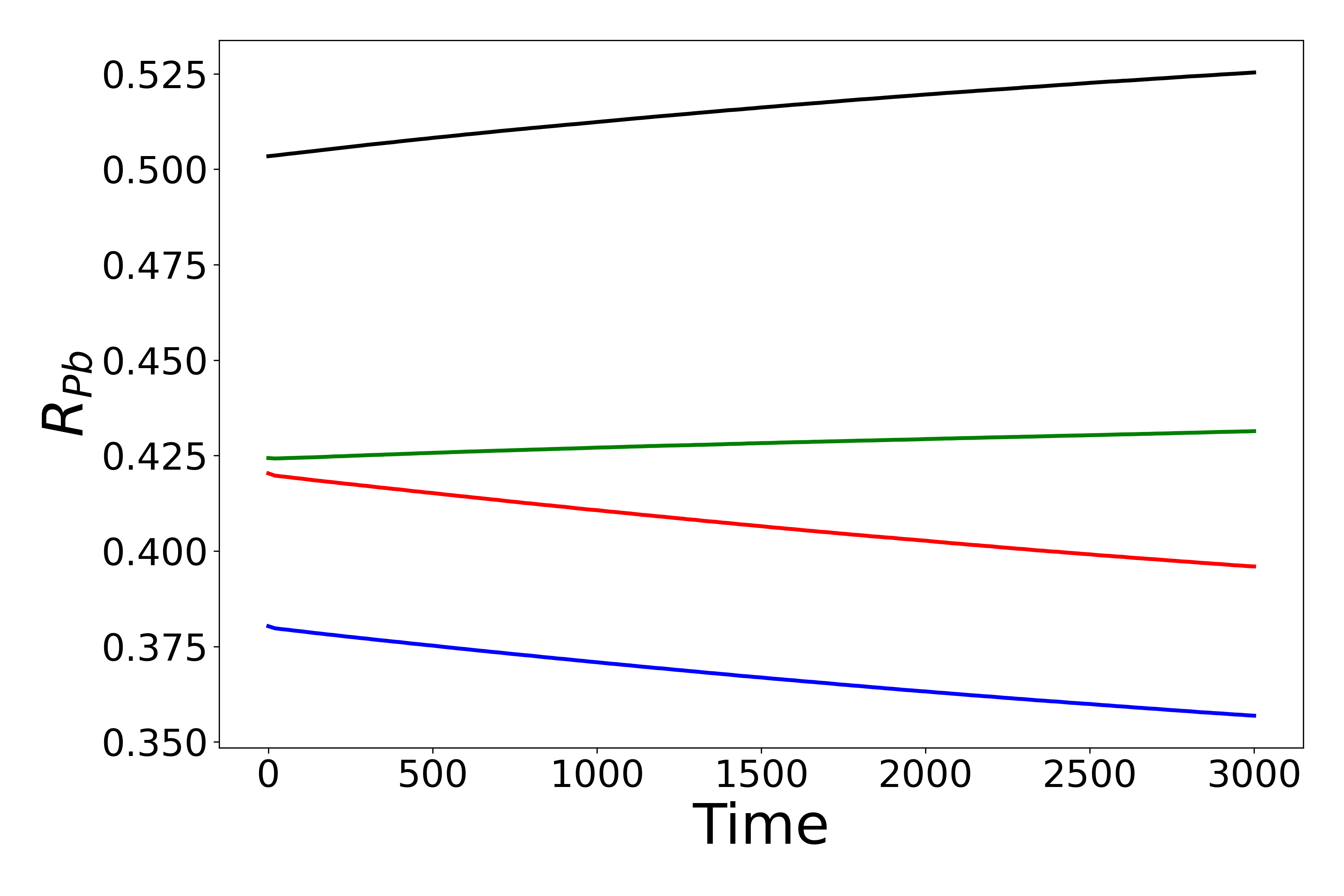 }
\includegraphics[width=0.32\linewidth]{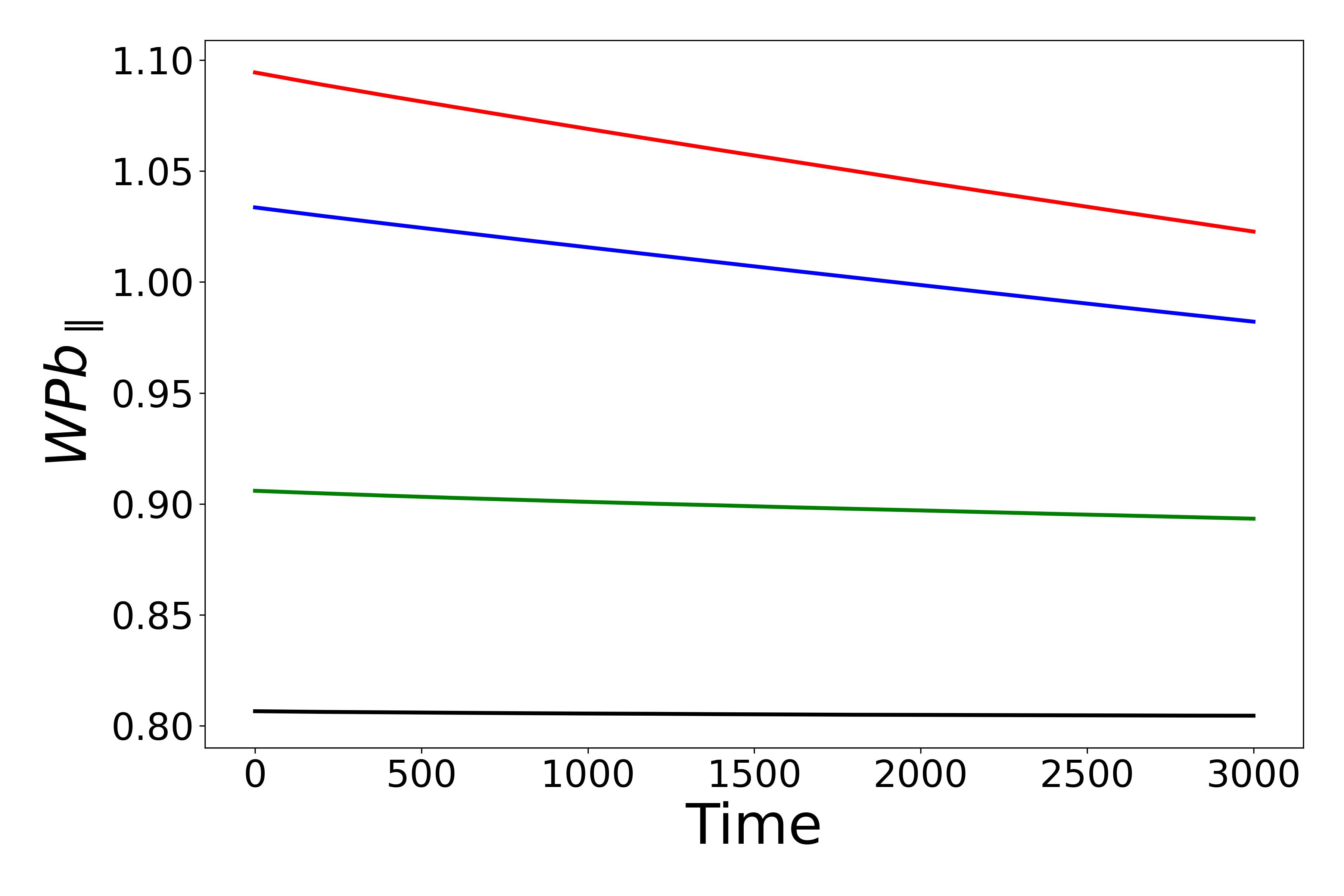}
\includegraphics[width=0.32\linewidth]{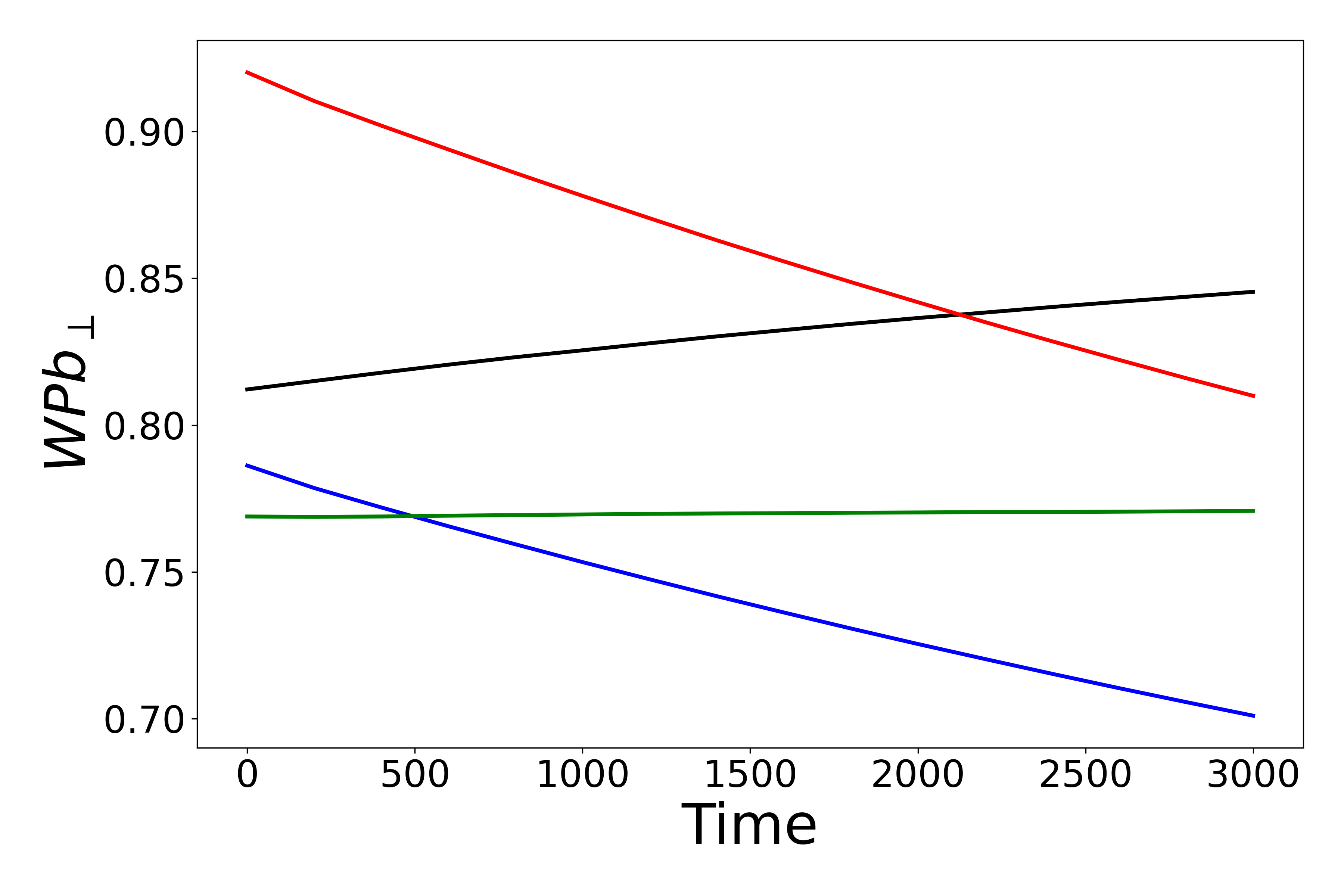}\\
(g)\hspace{2in}(h)\hspace{2in}(i)\\
\includegraphics[width=0.32\linewidth]{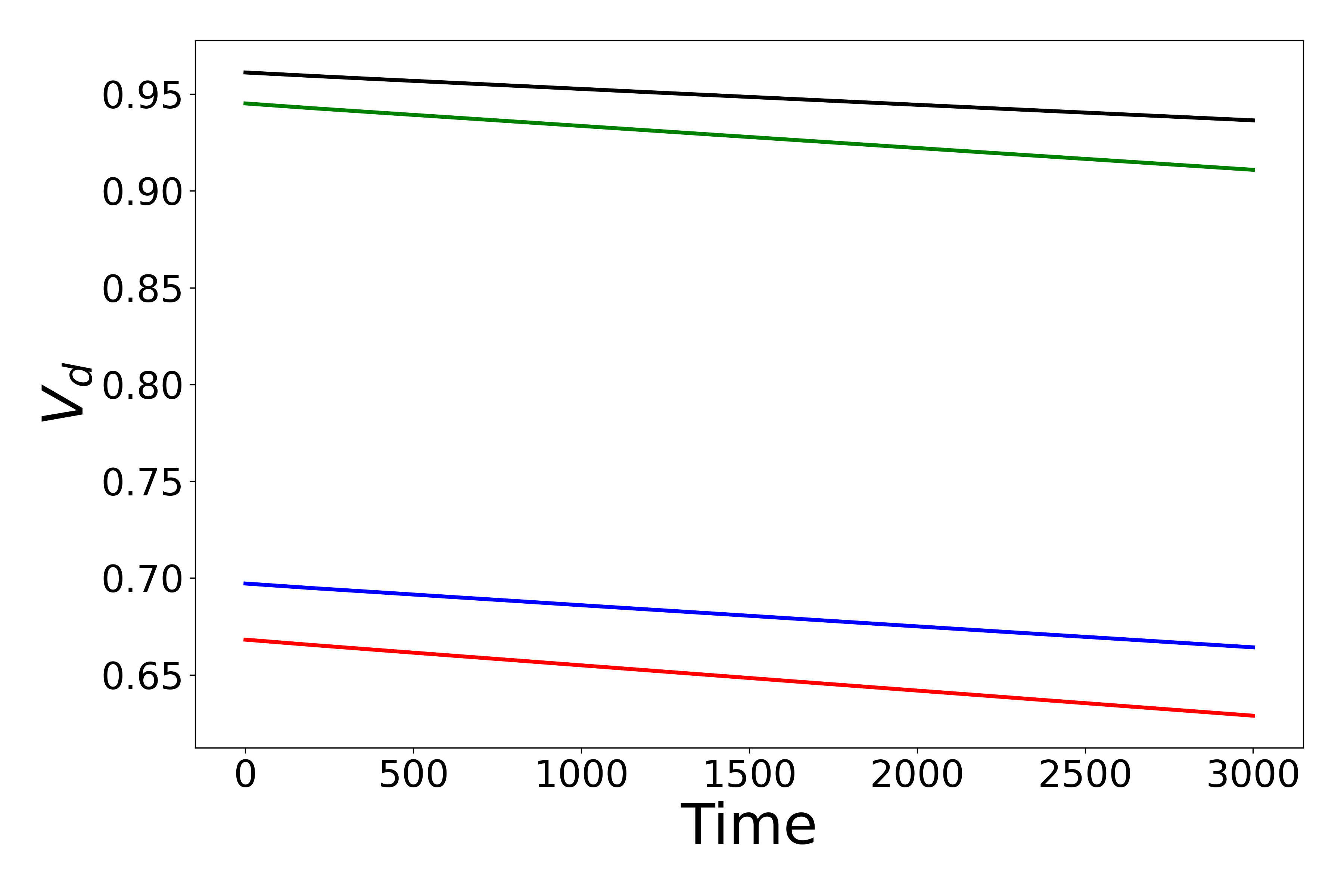}
\includegraphics[width=0.32\linewidth]{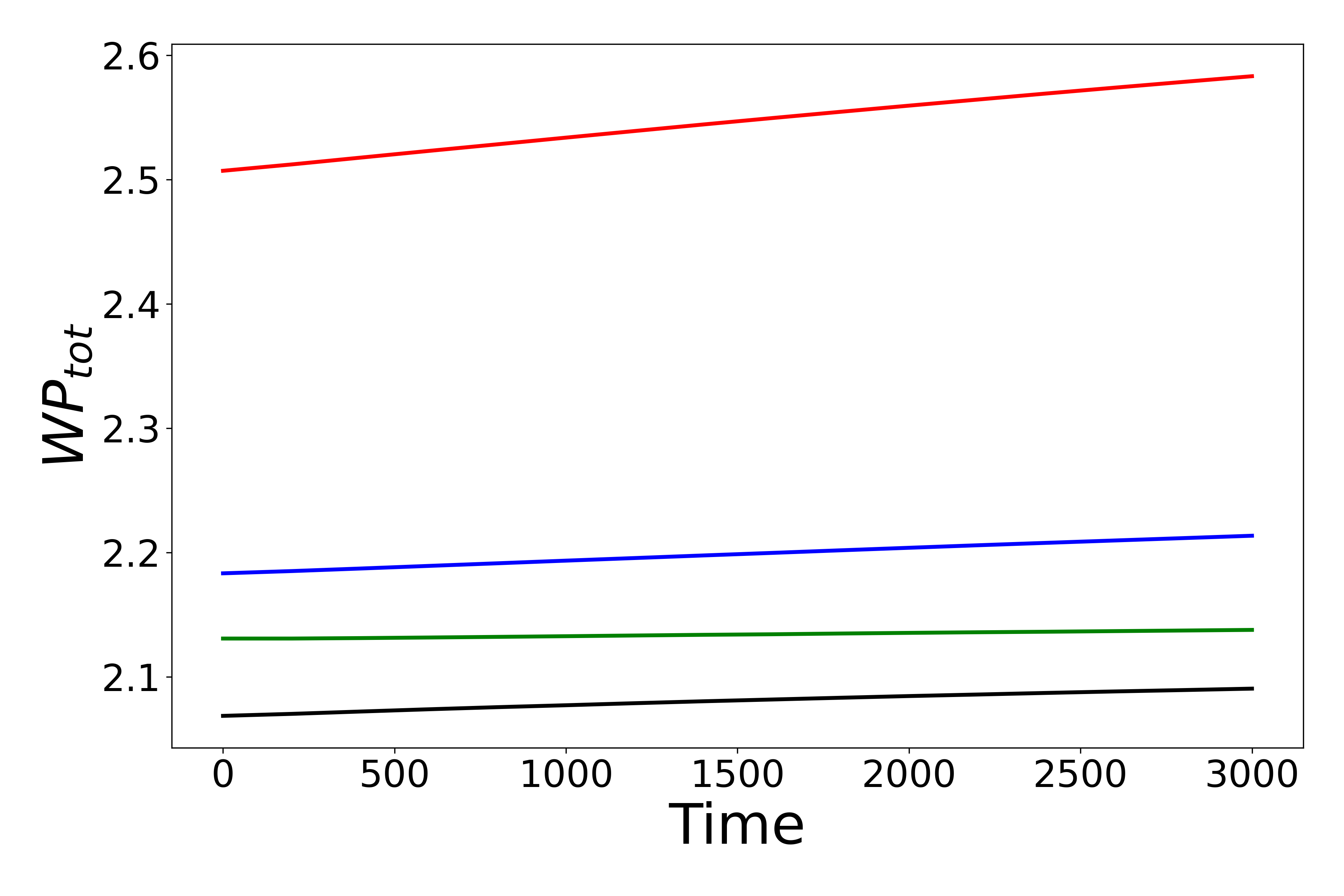}
\includegraphics[width=0.32\linewidth]{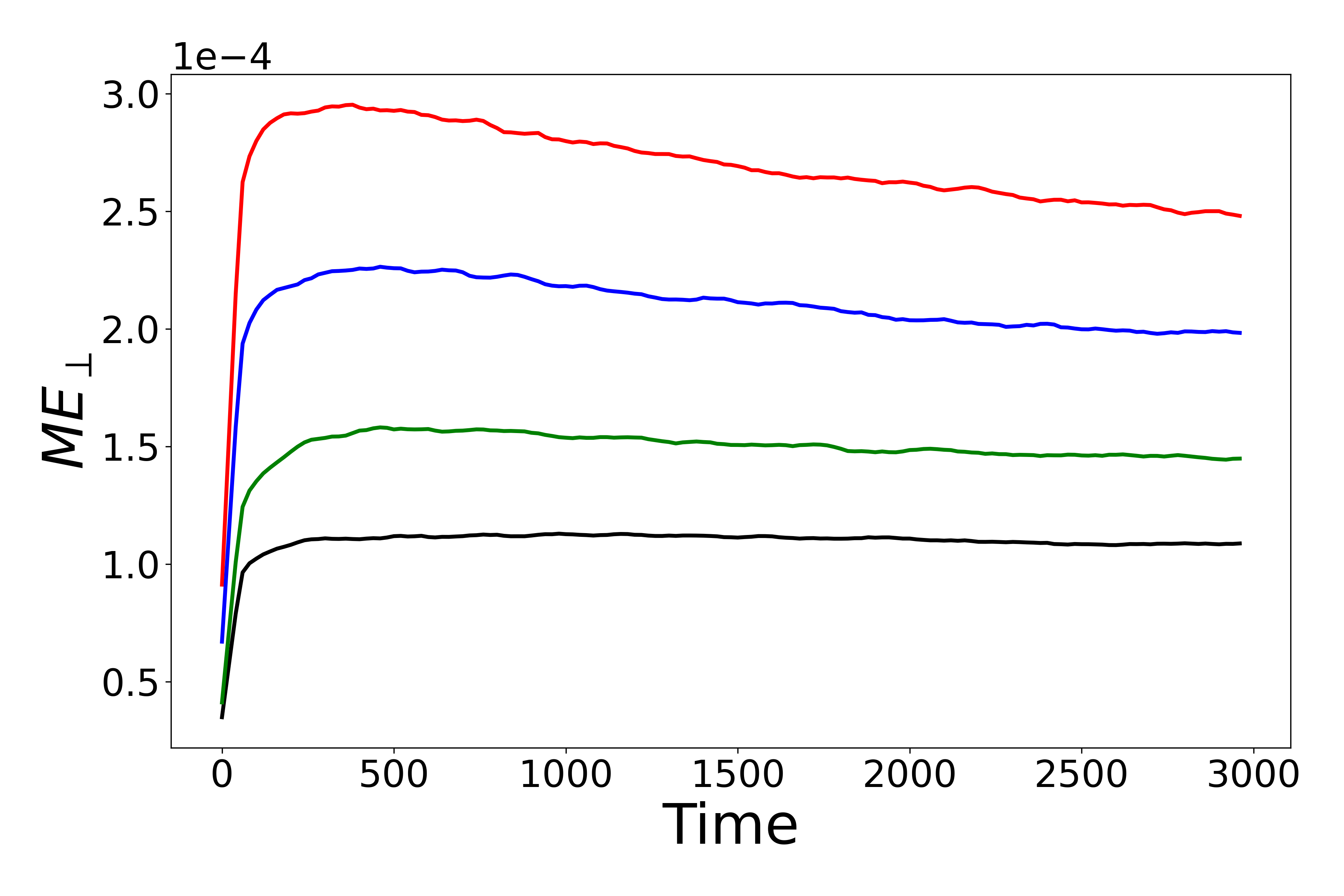}
\caption{Panels (a) to (i) show the hybrid model outputs for the cases listed in Table \ref{model_param:tab}. Panels (a) to (c) depict the variations in temperature anisotropy ($R_{PC}$), parallel thermal energy ($W_{PC_\parallel}$), and perpendicular thermal energy ($W_{PC_\perp}$) for proton cores. Panels (d) to (f) illustrate the variations in temperature anisotropy ($R_{Pb}$), parallel thermal energy ($W_{Pb_\parallel}$), and perpendicular thermal energy ($W_{Pb_\perp}$) for proton beams. Panels (g) to (i) show the drift velocity ($V_d$), the total thermal energy of proton ($WP_{tot}$) and perpendicular magnetic energy ($ME_\perp \sim B_\perp ^2$). Time is measured in proton gyroperiods. The initial beam-to-core number density ratio (bc) and drift velocity (vd) are also provided as the labels for each case. The roman numbers in label in panel (a) represent the class of each case.}

\label{fig:model}
\end{figure}

Figure~\ref{fig:model} illustrates the evolution of ion heating in both the parallel and perpendicular directions relative to the background magnetic field during the development and relaxation of kinetic instabilities. Panels (a), (b), and (c) of Figure \ref{fig:model} present the evolution of temperature anisotropy ($R_{PC}$) and the parallel ($W_{PC_\parallel}$) and perpendicular ($W_{PC_\perp}$) thermal energy of the proton core. Similarly, panels (d), (e), and (f) depict the same parameters—temperature anisotropy ($R_{Pb}$), and parallel ($W_{Pb_\parallel}$) and perpendicular ($W_{Pb_\perp}$) thermal energy values for the proton beams.

For Class I model parameters, the core proton temperature anisotropy decreases while beam proton temperature anisotropy gradually increases. In contrast, Class II models exhibits the opposite trend, with an increase in core proton temperature anisotropy and a reduction in beam proton temperature anisotropy. Additionally, Class II models shows a more significant increase ($\sim$40-50\%) in the thermal energy of the core protons compared to Class I models. For the proton beams, Class II models also demonstrate reduction ($\sim$ 20\%) in thermal energy relative to Class I models.

It is important to note that the vertical axis scales in panels (b), (c), (e), and (f) are different for the same modeling time interval. The results clearly indicate that the proton core in both parallel and perpendicular directions experiences substantial heating, whereas the beam part experiences cooling in both the directions.

Panels (g), (h), and (i) in Figure \ref{fig:model} explore the energy transfer between thermal, magnetic and kinetic energy due to drift velocity and temperature anisotropy instability. Panel (g) illustrates the variation in drift velocity ($V_d$) between the core and beam, with changes in drift velocity associated with changes in large-scale kinetic energy (K.E. $\propto$ $V_d^2$). Panels (h) and (i) show the change in the total thermal energy of the proton and perpendicular magnetic energy.

In the hybrid model, the initial background magnetic field and the bi-Maxwellian VDFs or the core and beam proton populations are specified based on PSP observations, and waves are generated self-consistently due to kinetic plasma instabilities that arise and evolve non-linearly in time. The rapid short-duration initial increase in perpendicular magnetic energy during the initial stage $t>0$ reflects the generation of these waves, followed by their self-consistent evolution with the VDFs, which represents energy exchange between particles and waves. For presently modeled cases, the gradual change in drift velocity throughout the temporal evolution, and consequently the small change in kinetic energy, indicates that most of the energy exchange occurs between the random kinetic energy of the particles (i.e., thermal energy) and the magnetic energy of the waves. This modeling result is corroborated by panels (h) and (i).

The evolution of the thermal energy of the proton core and beam populations can be used to calculate the growth and damping rates of the associated instabilities. The temporal evolution of these energies are fitted with the exponential function to calculate the growth/damping rate.  Since the proton beams exhibit a temperature anisotropy of less than one and the model output indicates that the beams are releasing energy, this released energy contributes to the generation of fast mode waves. The non-linear growth rates calculated from the beam energy range from $10^{-5}$ to $10^{-6}$ $\Omega_p^{-1}$.  

Meanwhile, the proton core absorbs energy, leading to the damping of ICWs. The nonlinear damping rates calculated from the proton core range from $10^{-6}$ to $10^{-4}$ $\Omega_p^{-1}$. Additionally, cases where the beam density exceeds the core density exhibit damping rates that are an order of magnitude higher which can also be understood from the nonlinear evolution of the ion thermal energy. This strong damping suggests significant attenuation of ICWs by the resonant interaction with the proton VDFs. These findings are consistent with the results from linear stability analysis also indicating ICW damping. The linear Vlasov equations are used to calculate the linear growth or damping rates. These linear rates for the observed frequency band are then compared with the non linear hybrid model growth/damping rates. The growth and damping rates calculated using linear stability analysis are presented in Appendix \ref{sec:lgr}. The results from the hybrid model and linear stability analysis show qualitative agreement, increasing the robustness of our findings.

The key takeaway from Sections \ref{sec:ls} and \ref{sec:mod} is that the proton core population is the primary absorber of wave energy, while the beam can emit ion-scale waves. The damping of the waves in the core is more prominent when the beam number density exceeds the core number density. These model results support the observation presented in Figure \ref{fig:eve}.

\section{Discussion and Conclusion}\label{sec:dis}

The generation and dissipation of ion-scale waves due to non-equilibrium ion VDFs has been previously studied extensively. However, most previous works for solar wind plasma applications have focused on plasma dynamics with relatively low-density secondary ion populations. In this paper, we illustrate the absorption of ion-scale waves by plasma with multiple proton components, with the beam and core populations having comparable densities, motivated by observations of such a case by PSP/SPAN-I during Encounter 17.

The PSP observations in Figure \ref{fig:eve} reveal the presence of ion-scale waves, including both left- and right-hand circularly polarized modes. Notably, the width of the frequency band where these waves are observed varies consistently with changes in the density ratio between proton cores and beams. Specifically, the frequency band becomes narrower when the number density of proton beams exceeds that of the proton cores. This dependence suggests the possibility of the ion-scale wave absorption by the protons and resonant heating.

To investigate this phenomenon further, we utilized the linear dispersion solver \texttt{PLUME} and the nonlinear 2.5D hybrid simulation model. \texttt{PLUME} provides the damping rates of the waves at present state in the linear approximation, while the hybrid model directly demonstrate nonlinear evolution of particle VDFs. Together, these tools offer a comprehensive picture of how ion-scale wave and proton velocity distributions in the core and beam populations interact.  

The \texttt{PLUME} solution indicates that the proton heating rate decreases as ion-scale wave energy diminishes. The proton core experiences greater heating during intervals when the beam to core density ratio is low, whereas the heating becomes more balanced between the two populations at the times when the beam to core density ratio is greater than unity.  The nonlinear hybrid model further reveals that during evolution, the proton core VDF absorb energy and cause strong damping of the ion-scale waves when the beam number density exceeds that of the core. The temperature anisotropies of the core and beam VDFs show contrasting behavior between distributions with low and high beam density relative to the core. The hybrid model shows minimal changes in non-thermal kinetic energy (drift velocity) over the long duration of three thousand proton gyroperiods (corresponding to $\sim$ 1028 seconds using the PSP observational parameters at E17 ). 

Although the growth and damping rates of the wave modes are small, they remain consistent between the nonlinear hybrid model and the linear stability analysis. Both approaches indicate that ICWs are damped by protons, with stronger damping observed when the proton beam density is comparable to the proton core density. This may explain the smaller wave band magnitude observed during periods when the beam density exceeds the core density.  

Our results enhance the understanding of ion-scale wave-particle interactions in the solar wind. The resonant absorption of waves could account for the reduced detection of ion-scale waves at larger distances from the Sun, in particular near the current sheet. This phenomenon warrants further investigation through statistical studies of wave-particle interactions, which will be addressed in future work.

\section*{Acknowledgment} 
We thank the PSP mission team for generating the data and making them publicly available. Y., L.O., S.A.B., and V.S. acknowledge support by NSF grant AGS-2300961. Y., L.O., S.A.B.,V.S., K.G.K., M.M., L.J., and P.M. acknowledge support by NASA grant  80NSSC24K0724. Resources supporting this work were provided by the NASA High-End Computing (HEC) Program through the NASA Advanced Supercomputing (NAS) Division at Ames Research Center. 
P. M. acknowledges the partial support by NASA HGIO grant
80NSSC23K0419 and the NSF SHINE grant 2401162. J.H. is supported by NASA grant 80NSSC23K0737.

Parker Solar Probe was designed, built, and is now operated by the Johns Hopkins Applied Physics Laboratory as part of NASA's Living with a Star (LWS) program (contract NNN06AA01C). Support from the LWS management and technical team has played a critical role in the success of the Parker Solar Probe mission. We also thank SPDF for making the data publicly available.
The PLUME code is publicly available at \url{https://github.com/kgklein/PLUME}.

\vspace{5mm}
\facilities{PSP (SWEAP, FIELDS)}

\appendix
\section{Linear Growth Rate}\label{sec:lgr}
\begin{figure}[h!]
\centering
\includegraphics[width=0.7\linewidth]{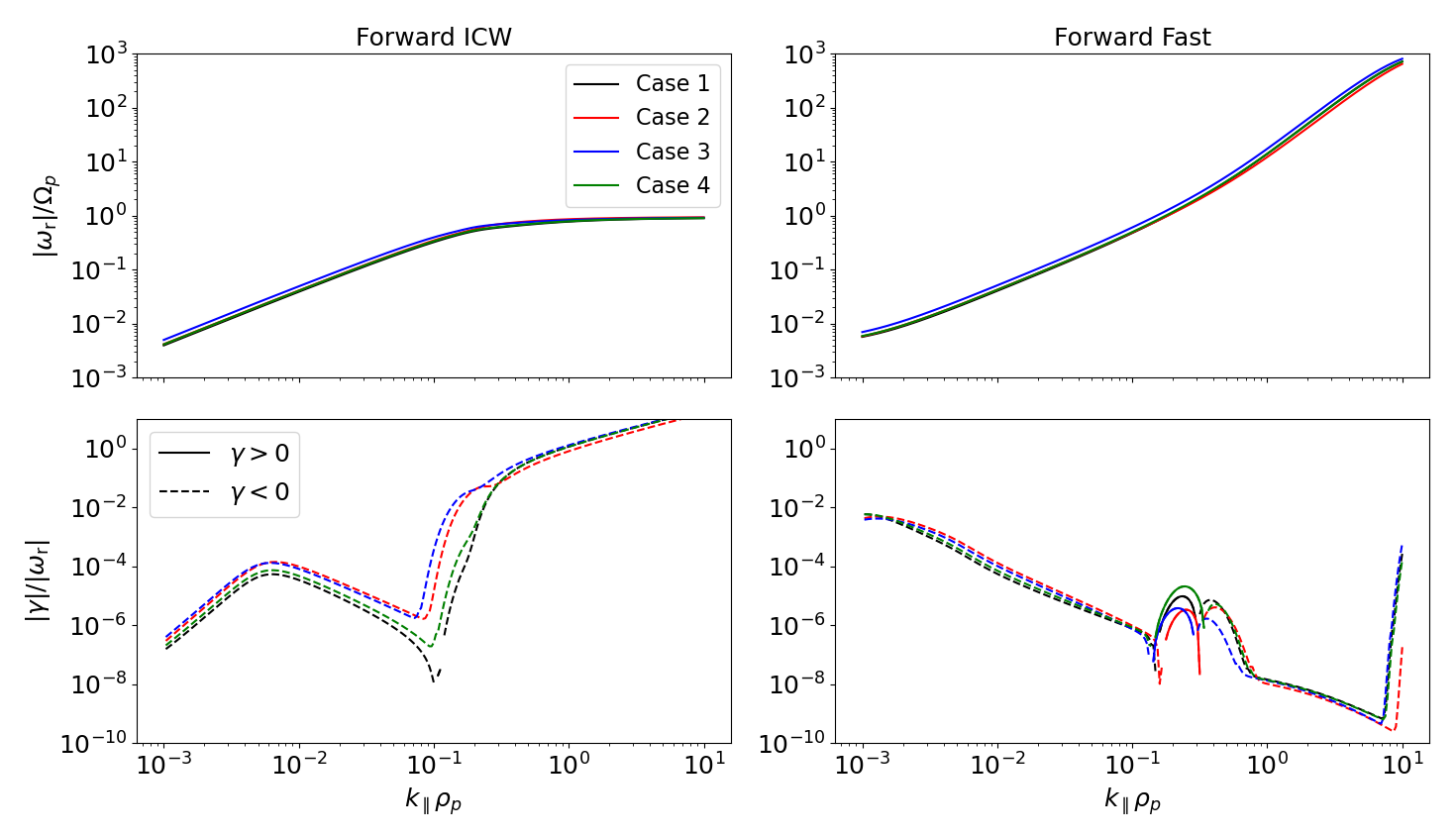}
\caption{Linear frequencies and growth/damping rates for the parallel forward Alfv\'{e}n/ICW (left column) and forward fast wave (right column). (Top row) Real frequency $\omega_{\textrm{r}}/\Omega_p$ and (Bottom row) growth/damping rate $|\gamma_j|/|\omega_{\textrm{r}}|$ for the designated solution for four selected cases. Solid and dashed lines represent positive and negative values respectively.}
    \label{fig:gr}
\end{figure}

Figure \ref{fig:gr} presents the linear frequencies and growth/damping rates for the forward parallel Alfv\'{e}n/Ion Cyclotron Wave and the forward fast wave mode. The input for the linear stability analysis are taken from Table \ref{model_param:tab}. The colors correspond to the cases shown in Figure \ref{fig:vdf}. The forward fast wave mode exhibits a small positive growth rate across all four cases. These fast waves can also be identified in Figure \ref{fig:eve} as the right-hand circularly polarized waves, shown in blue. The linear growth rate for fast modes varies between $10^{-4}$ and $10^{-6}$ $\Omega_p^{-1}$.

In contrast, the VDFs do not generate ICWs; instead, strong damping is observed. The damping rate is particularly high and ranges from $10^{-6}$ to $10^{-3}$ $\Omega_p^{-1}$ within the relevant frequency range (around 1 Hz). This rate is approximately one order of magnitude higher in cases where the beam density is comparable to the core density, compared to cases where the beam density is lower. To further investigate this behavior, we calculate the wave dissipation and the proton heating induced by these waves and the discussion can be seen in Section \ref{sec:ls}.


\bibliography{ref}{}
\bibliographystyle{aasjournal}

\end{document}